
\input harvmac

\let\preprint=0    


\def\PSfig#1#2#3#4#5{
  \nfig{#1}{#2}
  \if\preprint1
    \topinsert
      \dimen1=#3                     
      \divide\dimen1 by 100
      \multiply\dimen1 by #5
      \vskip \dimen1
      \count253=83                   
      \multiply\count253 #5
      \divide\count253 by 100
      \count254=100                  
      \advance\count254 by -#5
      \multiply\count254 by 254      
      \divide\count254 by 100
      \advance\count254 by -65       
      \dimen1=#3                     
      \advance\dimen1 by -11in
      \divide\dimen1 by 2
      \count255=\dimen1
      \divide\count255 by 100
      \multiply\count255 by \count253
      \dimen2=1bp
      \divide\count255 by \dimen2
      \advance\count255 by 18        
      \includegraphics{#4}
      \setbox1=\hbox{#1. #2}
      \ifdim \wd1 < 4.5in
         \centerline{#1. #2}
      \else
         \centerline{\vbox{\hsize 4.5in \baselineskip12pt
                     \noindent Fig.~\xfig#1. #2}}
      \fi
    \endinsert
  \fi
}


\def\refmark#1{${}^{\refs{#1}}$\ }
\def\footsym{*}\def\footsymbol{}\ftno=-2
\def\foot{\ifnum\ftno<\pageno\xdef\footsymbol{}\advance\ftno by1\relax
\ifnum\ftno=\pageno\if*\footsym\def\footsym{$^\dagger$}\else\def\footsym{*}\fi
\else\def\footsym{*}\fi\global\ftno=\pageno\fi
\xdef\footsymbol{\footsym\footsymbol}\footnote{\footsymbol}}

\font\authorfont=cmcsc10 \ifx\answ\bigans\else scaled\magstep1\fi
\edef\smalltfontsize{\ifx\answ\bigans scaled\magstep2\else scaled\magstep3\fi}
\font\smalltitlefont=cmr10 \smalltfontsize

\baselineskip=0.29in plus 2pt minus 2pt

\def\lsim{ \vcenter{\hbox{$\buildrel{\displaystyle <}\over\sim$}} }

\def\pt{p_\bot}
\def\kt{k_\bot}
\def\pz{p_{\rm z}}
\def\pza{p_{\rm 1z}}
\def\pzb{p_{\rm 2z}}
\def\vza{v_{\rm 1z}}
\def\vzb{v_{\rm 2z}}

\def\d{{\rm d}}

\def\Vcl{V_{\rm cl}}
\def\Veff{V_{\rm eff}}
\def\PiR{\Pi_{\rm R}}
\def\Pnet{P_{\rm net}}
\def\phicl{\phi_{\rm cl}}
\def\M{{\cal M}}
\def\Im{{\rm Im}}

\def\T{{\cal T}}
\def\R{{\cal R}}

\def\opr{{\mathop{\partial}\limits^{\leftrightarrow}}_z^2}

\lref\shaposhnikov{
  M. Shaposhnikov, JETP Lett.\ \bf 44\rm, 405 (1986);
    Nucl.\ Phys.\ \bf B299\rm, 707 (1988),
}
\lref\mclerrana{
  L. McLerran, Phys.\ Rev.\ Lett.\ \bf 62\rm, 1075 (1989).}
\lref\tzmsv{
  N. Turok and J. Zadrozny, Phys.\ Rev.\ Lett.\ \bf 65\rm, 2331 (1990);
    Nucl.\ Phys.\ \bf B358\rm, 471 (1991);
  L. McLerran, M. Shaposhnikov, N. Turok and M. Voloshin,
    Phys.\ Lett.\ \bf 256B\rm, 451 (1991).
}
\lref\cohen{
  A. Cohen, D. Kaplan and A. Nelson,
    Phys.\ Lett.\ \bf B245\rm, 561 (1990);
    Nucl.\ Phys.\ \bf B349\rm, 727 (1991);
    Phys.\ Lett.\ \bf B263\rm, 86 (1992).
}
\lref\dine{
  M. Dine, P. Huet, R. Singleton and L. Susskind,
    Phys.\ Lett.\ \bf B257\rm, 351 (1992);
  M. Dine, P. Huet and R. Singleton, Nucl.\ Phys.\ \bf B375\rm, 625 (1992).
}
\lref\review{
  For a good review, see A. Cohen, D. Kaplan and A. Nelson,
  UCSD-PTH-93-02, to appear in \sl Ann.\ Rev.\ Nucl.\ Part.\ Sci.\
  \bf 43\rm.}
\lref\dinea{
  M. Dine, R. Leigh, P. Huet, A. Linde and D. Linde,
    Phys.\ Rev.\ \bf D46\rm, 550 (1992);
    Phys.\ Lett.\ \bf B283\rm, 319 (1992).
}
\lref\liu{
  B-H.\ Liu, L. McLerran and N. Turok, Phys.\ Rev.\ \bf D46\rm, 2668 (1992).
}
\lref\khlebnikov{
  S. Khlebnikov, Phys.\ Rev.\ \bf D46\rm, 3223 (1992).
}
\lref\huet{
  P. Huet, K. Kajantie, R. Leigh, B-H.\ Liu and L. McLerran,
  SLAC-PUB-5943.
}
\lref\lindev{
  A. Linde, Nucl.\ Phys.\ \bf B216\rm, 421 (1983).
}
\lref\turokq{
  N. Turok, Phys.\ Rev.\ Lett.\ \bf 68\rm, 1803 ( 1992).
}
\lref\weldon{
  H.A. Weldon, Phys.\ Rev.\ \bf D28\rm, 2007 (1983).
}
\lref\scatter{
  J. Taylor, ``Scattering Theory'' (John Wiley \& Sons: New York, 1972).
}

\Title{UW/PT-93-2
  }{\vbox{
    \smalltitlefont
    \centerline{One-loop fluctuation-dissipation formula}
    \centerline{for bubble-wall velocity}
  }}
\centerline{\authorfont Peter Arnold}
\centerline{\sl Department of Physics, F-15, University of Washington,
    Seattle, WA  98195}

\vskip .1in
The limiting bubble wall velocity during a first-order electroweak
phase transition is of interest in scenarios for electroweak baryogenesis.
Khlebnikov has recently proposed an interesting method for computing
this velocity based on the fluctuation-dissipation theorem.
I demonstrate that at one-loop order this method is identical to simple,
earlier techniques for computing the wall velocity based on computing
the friction from particles reflecting off or transmitting
through the wall in the ideal gas limit.

\Date{February 1993}

There has recently been considerable interest in the limiting bubble
wall velocity during a first-order electroweak phase
transition.\refmark{\dinea\liu\khlebnikov-\huet}
The dynamics of the bubble wall bears on scenarios for electroweak
baryogenesis mediated by the anomalous violation of baryon number in the
Standard Model.\refmark{\shaposhnikov\mclerrana\tzmsv\cohen\dine-\review}
During the phase transition, bubbles of the
symmetry-broken phase are nucleated inside the symmetric phase, and
these bubbles then expand to convert the entire Universe into the
symmetry-broken phase.  At zero temperature, in vacuum, the velocity of the
bubble walls would asymptotically approach the speed of light; at finite
temperature, however, collisions of particles off of the bubble wall
provide a source of friction and force a smaller limiting velocity.
A simple model for this friction is to compute the momentum transfer
to the wall from an ideal gas of thermally-distributed particles
colliding with, and either reflecting from or transmitting through,
the bubble wall.\foot{
  This has been discussed, in various limits, in
  refs.~\refs{\dinea,\liu,\lindev,\turokq}.
}
This model is valid provided the thermal mean free
path of those particles is large compared to the thickness of the
bubble wall.  (This is not a particularly good assumption for the real
case of the electroweak phase transition, but is often used to
establish a lower limit on the bubble wall velocity.)
Khlebnikov\refmark{\khlebnikov} has recently
proposed an interesting alternative method for computing the bubble wall
velocity $u$,
based on the fluctuation-dissipation theorem, for cases where $u \ll 1$.
When he puts his general formalism into practice with a sample one-loop
calculation, he claims to find slightly different results than expected
from other methods.
In this paper, however, I shall show that the one-loop
result obtained from the fluctuation-dissipation theorem is identical
to the simple model of reflection and transmission in the limit of
small wall velocities and large mean free path.  (Khlebnikov's
general formalism is still of interest because it applies to
higher-loop calculations as well, always provided $u \ll 1$.
For example, the effects of rescattering, which go beyond the approximations
considered in this paper, originate in higher-order loops.)

I shall also reexamine a toy model calculation considered
by Khlebnikov, where he found a logarithmic suppression of the bubble
wall velocity.  It will be possible to see the physical origin of
this logarithm (and of another logarithm that he accidentally omitted)
by using the picture of reflection and transmission at the bubble
wall.

\newsec{Review of simple reflection/transmission model}

\PSfig\figa{Qualitative form of the effective, finite-temperature
  Higgs potential during a first-order electroweak
  phase transition.}{10.5cm}{fig1.psx}{60}

\PSfig\figb{Qualitative profile of the bubble wall.}{4cm}{fig2.psx}{75}

Figure~\xfig\figa\ shows the qualitative form of the
effective, finite-temperature Higgs potential
when the phase transition begins: the symmetric $\phi=0$ phase (phase 1)
is unstable to decay into the asymmetric vacuum (phase 2).
Figure~\xfig\figb\ shows
qualitatively the profile of a bubble wall interpolating between the
phases and which, as drawn, will move from right to left.  Since
particle masses depend on the value of $\phi$, masses will be
different in the two phases.

If the mean free path is large compared to the wall thickness, then we
may separately treat the collision of each particle in the plasma
with the bubble wall.  In the wall rest frame, the wall will be struck
by a Lorentz-boosted thermal distribution of particles.  I shall focus
on the case of bosons.
The usual Bose distribution $n_p = 1/[\exp(\beta E_p)-1]$ is the
form, in the thermal rest frame, of the Lorentz
scalar $1/[\exp(\beta p \cdot v) - 1]$ where $v$ is the four-velocity
of the thermal bath.  In the rest frame of the bubble wall, moving
with wall velocity $u$ relative to the thermal bath, this distribution
is
\eqn\nboost{
   f_p =
   { 1 \over \exp[\beta\gamma(E_p+\vec p\cdot \vec u)] - 1 } .
}
Take $\vec u$ to be in the $z$ direction, and
consider a particle of energy $E$ and momentum $\vec p = (\vec\pt,\pz)$
incident from the symmetric phase.  Both $E$ and $\vec\pt$ will be conserved.
The momentum transferred to the wall will be $2\pza$ if the particle is
reflected and $\pza-\pzb$ if it is transmitted, where
\eqn\eqpzab{
  \pza = \sqrt{E^2-\pt^2-m_1^2},
  \qquad
  \pzb = \sqrt{E^2-\pt^2-m_2^2},
}
and $m_1$ and $m_2$ are the masses of the particle in the two phases.
The pressure on the wall from particles incident from the symmetric phase
is then
\eqn\Peqna{
   P_1 = \int {\d^3 p_1\over(2\pi)^3} \vza \theta(\vza) f_{p1}
     \left[ 2 \pza \R(p_1)
     + (\pza-\pzb) \T(p_1) \right] ,
}
where $\R(p_1)$ and $\T(p_1)$ are the reflection and transmission coefficients
for particles encountering the wall
and $\vec v_1$ is the incident particle velocity $\vec p_1/E$.
The pressure $P_2$ from particles
incident from the asymmetric phase is similar.\foot{
  I shall assume throughout that there the temperature does not vary
  appreciably across the wall.  This is the case for the electroweak phase
  transition; see refs.~\dinea\ and \turokq.
}
Also, the difference
of the Higgs potential between the two phases exerts pressure on the
wall.  The net pressure is
\eqn\eqPnet{
   \Pnet = P_1 - P_2 - \Delta\Vcl ,
}
where $\Vcl$ is the classical, zero-temperature Higgs potential.
(The finite-temperature effective potential will appear momentarily.)

Consider this result in the limit of zero wall velocity $u$, and change
integration variables in \Peqna\ from $\pz$ to $E$:
\eqn\pZeroU{
  \eqalign{
   \Pnet(u\!=\!0) =
   \int {\d^2 p_\bot \, \d E\over(2\pi)^3} & {1\over e^{\beta E}-1}
   \bigl\{
     2 \pza \theta (-\pzb^2)
   \cr&
     + 2(\pza - \pzb) [ \R(\pt,E) + \T(\pt,E) ] \theta(\pzb^2)
   \bigr\} - \Delta\Vcl,
  }
}
where I have used the fact from one-dimensional quantum mechanics that
reflection and transmission coefficients do not depend from which side
the particle approaches.  $\theta(\pm\pzb^2)$ above is a short-hand
notation for $\theta[\pm(E^2-\pt^2-m_2^2)]$, which specifies whether or
not the particle is classically allowed in phase 2.
The first term in braces above corresponds to particles from phase 1
which are not energetic enough to enter phase 2 and so must reflect
from the bubble wall; the second term corresponds to more energetic
particles, which may come from phase 1 or phase 2 and may reflect or
transmit.
Now use $\R+\T=1$ and integrate by parts to get
\eqn\pZeroUb{
  \Pnet(u=0) =
  T \int {\d^3 p\over(2\pi)^3} \left[
    \ln\left(1-e^{-\beta E_1}\right) - \ln\left(1-e^{-\beta E_2}\right)
  \right] - \Delta \Vcl,
}
which is simply the difference $-\Delta\Veff$ in the one-loop
finite-temperature effective potential between the two
phases.\refmark{\turokq}

At {\it non}-zero wall velocity, the net pressure \Peqna\
may then be written as
\eqn\Pmoo{
  \eqalign{
   \Pnet = &
     \int {\d^3 p_1\over(2\pi)^3} \vza \theta(\vza) \Delta f_{p1}
       \left[ 2 \pza \R(p_1) + (\pza-\pzb) \T(p_1) \right]
   \cr&
     -
     \int {\d^3 p_2\over(2\pi)^3} \vzb \theta(-\vzb) \Delta f_{p2}
       \left[ 2 \pzb \R(p_1) + (\pzb-\pza) \T(p_1) \right]
     - \Delta\Veff,
  }
}
where $\Delta f_p = f_p - n_p$.  In the limit of small wall
velocity $u$, $f_p - n_p \approx u\beta\pz n_p (n_p+1)$, and
\Pmoo\ simplifies to
\eqn\Psmall{
  \eqalign{
   \Pnet =
     - \Delta\Veff + \beta u \int & {\d^2 \pt \d E \over (2\pi)^3}
     n(E) [n(E)+1] \bigl\{ 2\pza^2 \theta(-\pzb^2)
  \cr&
       + \left[ 2(\pza^2+\pzb^2) \R(\pt,E) + (\pza-\pzb)^2 \T(\pt,E) \right]
           \theta(\pzb^2)
     \bigr\} + O(u^2),
  }
}
where $\pza$ and $\pzb$ will henceforth always be understood to be positive,
according to \eqpzab.
The limiting velocity of the wall is determined by setting $\Pnet = 0$.

\PSfig\figc{Schematic depiction of two solutions to the Schr\"odinger
  equation describing a particle propagating in the background of the
  wall.  A wave may be incident from either (a) the left or (b) the
  right, and it reflects and transmits at the wall.  The heavy
  solid line depicts the potential.}{4cm}{fig3.psx}{80}

Equation \Psmall\ is the formula that I shall show is
equivalent to the one-loop
fluctuation-dissipation result of Khlebnikov.\refmark{\khlebnikov}
For the sake of simplicity
and concreteness,
I shall work with a generalization of the toy scalar model
considered by Khlebnikov.
The Lagrangian is
\eqn\eqLtoy{
   {\cal L} = {1\over2} (\partial\phi)^2 - V(\phi)
              + {1\over2}(\partial\Phi)^2
              - {1\over2} M^2(\phi) \Phi^2 ,
}
where
$\phi$ is the usual Higgs field, and $\Phi$ is another particle in the
theory that, analogous to W and Z bosons, has a mass that depends on
the VEV of $\phi$.  [Khlebnikov's specific model was
$M^2(\phi) = M_0^2 + 2g\phi$.]
In the presence of a background field $\phicl(z)$
describing the bubble wall in its rest frame, the $\Phi$ field has
the effective Lagrangian
\eqn\LPeff{
   {\cal L}_{\rm eff} =
   {1\over2}(\partial\Phi)^2 - {1\over2} M^2(\phicl(z)) \Phi^2 .
}
This is just the problem of a particle propagating in a classical potential.
The S matrix may be found by solving the classical equation of motion,
taking a solution with a single incoming wave, and extracting the
amplitudes of outgoing waves.  Fourier transforming $(t,\vec x_\bot)$
to $(E,\vec\pt)$ and noting that the latter are conserved, the equation
of motion is just
\eqn\Leqm{
    \left[ - \partial_z^2 + M^2(\phicl(z)) \right] \Phi
    = (E^2-\pt^2) \Phi ,
}
which is nothing more than the Schr\"odinger equation for a unit mass
particle in the potential $2 M^2(\phicl(z))$ with energy
$2(E^2-\pt^2)$.  For incident waves from the right or left, it
has solutions shown qualitatively in \figc, and the reflection
and transmission coefficients extracted from these solutions are
the $\R$ and $\T$ which appear in \Psmall.

\newsec{Fluctuation-dissipation theorem at one loop.}

I shall now show that one also obtains the result \Psmall\ from the
fluctuation-dissipation theorem at one-loop.
Khlebnikov treats the
difference in vacuum energies in \figa\ as a small perturbation.
Imagine turning off that perturbation, so that the vacua are degenerate.
A bubble wall in this situation would then remain stationary.
Now turn the perturbation back on.  The non-equilibrium response
of the system to this perturbation can then be related, via the
fluctuation-dissipation theorem, to the {\it equilibrium} calculation
of a retarded correlator.  Specifically, assuming the bubble wall
velocity is small, he finds
\eqn\fluct{
  u = \Delta\Veff
  \lim_{\omega\rightarrow0}
  \left[ {1\over\omega} \int \d z \d z' \,
    \partial_z \phicl(z) \,
    {\rm Im} \PiR(\omega;\vec\kt\!=\!0; z,z') \,
    \partial_{z'} \phicl(z')
  \right]^{-1}
  ,
}
where $\PiR$ is the retarded self-energy of the Higgs field $\phi$.
$\PiR$ is projected above onto the translational zero-mode
$\partial_z \phicl(z)$ of the bubble wall.  In position space,
$\PiR = {\langle[\phi(r),\phi(r')]\rangle}$
is a function of two space-time coordinates.
I have Fourier transformed the $t$, $x$, and $y$ components above
to $\omega$ and $\vec\kt$ but have left the $z$ coordinates in position space.
The trick is to now compute the desired limit of $\PiR$ in
the {\it background} of the
bubble wall $\phicl(z)$.
I shall examine this problem at one-loop order.

\PSfig\figd{One-loop contributions of $\Phi$ to the
  Higgs self-energy.}{5.5cm}{fig4.psx}{80}

For simplicity, return to the scalar model \eqLtoy\ and focus on
the one-loop contributions to $\PiR$ from $\Phi$, which are shown
in \figd.  Only \figd b has an imaginary part.
To get its contribution to $\Im\PiR$, first consider
evaluating the self-energy in Euclidean space for Euclidean frequencies
$i\nu$.  Figure~\xfig\figd b gives
\eqn\pia{
  \eqalign{
    \Pi(i\nu;0;z,z') =
    {1\over2} \M'(z) & \M'(z') T \sum_{p_0} \int {\d^2\pt\over(2\pi)^2}
  \cr& \times
    \left(1\over p_0^2-\pt^2-\Delta^2\right)_{z z'}
    \left(1\over (p_0+i\nu)^2 - \pt^2 - \Delta^2\right)_{z' z} ,
  }
}
where the sum is over discrete $p_0 = i2\pi nT$,
$\Delta^2$ is the operator
\eqn\eqDelta{
  \Delta^2 = -\partial_z^2 + M^2(\phicl(z)) ,
}
and
\eqn\MMdef{
  \M'(z) = \partial_\phi M^2(\phicl(z)) .
}
Now let $f_\kappa^\pm(z)$ be the eigenfunctions of $\Delta^2$ defined
by
\eqn\eigen{
   \Delta^2 f_\kappa^\pm = \kappa^2 f_\kappa^\pm
}
and normalized to
\eqn\norm{
   \int \d z \, {f_\kappa^a}^*(z) f_{\kappa'}^b(z)
   = \delta^{ab} \delta(\kappa^2 - \kappa'^2) .
}
Later on, we shall see that a useful basis for the $f^\pm$ will be
the solutions depicted in \figc, representing scattering of
particles incident from a definite side of the wall.
By inserting a complete set of states, $\pia$ may be rewritten as
\eqn\pib{
  \eqalign{
    \Pi(i\nu;0;z,z') =
    {1\over2} \M'(z) & \M'(z') \int \d\kappa^2 \d\kappa'^2
    T \sum_{p_0} \int {\d^2\pt\over(2\pi)^2}
  \cr& \times
    {f_\kappa^a}^*(z) {1\over p_0^2-\pt^2-\kappa^2} f_\kappa^a(z')
    {f_{\kappa'}^b}^*(z') {1\over (p_0+i\nu)^2-\pt^2-\kappa'^2}
      f_{\kappa'}^b(z) ,
  }
}
where sums over $a,b=\pm$ are implicit.
At this point, the steps to obtain $\Im\PiR$ are exactly the same as
they would be in the case of {\it free} scalar propagators and may be
found, for example, in refs.~\refs{\weldon,\khlebnikov}.
First the frequency sum over $p_0$
is performed by the usual contour trick to obtain
\eqn\pic{
  \eqalign{
    \Pi(i\nu;0;z,z') =
    {1\over2} \M'(z) & \M'(z') \int \d\kappa^2 \d\kappa'^2
    {f_\kappa^a}^*(z) f_{\kappa'}^b(z) {f_{\kappa'}^b}^*(z') f_\kappa^a(z')
    \int {\d^2\pt\over(2\pi)^2}
  \cr \times
    \biggl\{ &
      {n(E)\over 2E}   \biggl[ {1\over (E+i\nu)^2  - (\pt^2+\kappa'^2)}
                             + {1\over (E-i\nu)^2  - (\pt^2+\kappa'^2)} \biggr]
  \cr
    + & {n(E')\over 2E'} \biggl[ {1\over (E'+i\nu)^2 - (\pt^2+\kappa^2)}
                             + {1\over (E'-i\nu)^2 - (\pt^2+\kappa^2)} \biggr]
    \biggr\}
  \cr
    + & ~ \hbox{(T-independent terms)} ,
  }
}
where
\eqn\Edefk{
  E = \sqrt{\pt^2+\kappa^2}, \qquad E' = \sqrt{\pt^2+\kappa'^2} .
}
The retarded self-energy is then obtained by replacing $i\nu$ by
$\omega-i\epsilon$.  It is now straightforward to take the
imaginary part,
\eqn\picc{
  \eqalign{
    \Im\PiR(\omega;0;z,z') =
    {1\over2} \M'(z) & \M'(z') \int \d\kappa^2 \d\kappa'^2
    {f_\kappa^a}^*(z) f_{\kappa'}^b(z) {f_{\kappa'}^b}^*(z') f_\kappa^a(z')
    \int {\d^2\pt\over(2\pi)^2}
  \cr & \times
    {\pi\over 4E E'} [n(E)-n(E')]
    [\delta(\omega+E-E') - \delta(\omega-E+E')] ,
  }
}
and to take the $\omega\rightarrow 0$ limit.  Switching integration
variables from $\kappa$ to $E$,
\eqn\pid{
  \eqalign{
    \Im\PiR(\omega;0;z,z') \rightarrow
    \pi\beta\omega \M'(z) & \M'(z') \int {\d^2\pt\over(2\pi)^2}
    \int\d E
  \cr& \times
    n(E) [n(E)+1]
    {f_\kappa^a}^*(z) f_\kappa^b(z) {f_\kappa^b}^*(z') f_\kappa^a(z').
  }
}
Projecting this onto the bubble wall zero mode then transforms Khlebnikov's
result \fluct\ into:
\eqn\pie{
  {\Delta\Veff\over u} = \pi\beta \int {\d^2\pt\over(2\pi)^2} \d E\,
     n(E) [n(E)+1] \sum_{a,b}
     \left| \int\d z\, {f_\kappa^a}^*(z) \partial_z M^2(\phicl(z))
        f_\kappa^b(z) \right|^2 .
}
This will reproduce the result from \Psmall\ provided the sum over
$(a,b)$ above can be appropriately related to the reflection and
transmission coefficients $\R$ and $\T$.

Using the eigenequation \eigen\ that defines the $f_\kappa^a$, it is
straightforward to show that the $z$-integrand is a total derivative:
\eqn\parts{
   {f_\kappa^a}^*(z) \partial_z M^2(\phicl(z)) f_\kappa^b(z)
   = \partial_z \left[ {1\over2} {f_\kappa^a}^* \opr f_\kappa^b\right] ,
}
where $\opr$ is defined by
\eqn\oprdef{
   f \opr g \equiv (\partial_z^2 f) g - 2 (\partial_z f) (\partial_z g)
              + f (\partial_z^2 g) .
}
So the result \pie\ depends only on the values of
${f_\kappa^a}^*\opr f_\kappa^b$ at infinity,
where $f_\kappa^a$
becomes a superposition of incoming and outgoing plane waves.
On plane waves, note that $\opr$ gives
\eqn\oprplane{
   \left( e^{ikz} \right)^* \opr e^{ikz} = - 4 k^2 ,
   \qquad
   \left( e^{-ikz} \right)^* \opr e^{ikz} = 0 .
}
Now let us choose the basis for $f_\kappa^\pm$ depicted in \figc\ and
parameterize the asymptotic behavior as
\eqn\fbasis{
  \eqalign{
    f_\kappa^{+}(z) \rightarrow
    &\cases{
       {1\over\sqrt{4 \pi k_1}} \left( e^{ik_1z} + S_{++} e^{-ik_1z} \right) ,
         & $x \rightarrow -\infty$ , \cr
       {1\over\sqrt{4 \pi k_2}} S_{+-} e^{ik_2z} ,
         & $x \rightarrow +\infty$ ,
    }
  \cr
    f_\kappa^{-}(z) \rightarrow
    &\cases{
       {1\over\sqrt{4 \pi k_1}} S_{-+} e^{-ik_1z} ,
         & $x \rightarrow -\infty$ , \cr
       {1\over\sqrt{4 \pi k_2}} \left( e^{-ik_2z} + S_{--} e^{ik_2z} \right) ,
         & $x \rightarrow +\infty$ ,
    }
  }
}
where the $S$ are functions of $\kappa$ and
\eqn\kapdefb{
  \kappa^2 = k_1^2 + M^2(-\infty) = k_2^2 + M^2(+\infty) .
}
The matrix of coefficients
\eqn\Sdef{
  S = \pmatrix{ S_{++} & S_{+-} \cr S_{-+} & S_{--} }
}
is symmetric and unitary.\foot{
  This follows from examining the Wronskians $W$ of $f^+$ with $f^-$ or
  ${f^-}^*$ and equating \hbox{$W(z{=}-\infty)$} with
  \hbox{$W(z{=}+\infty)$}.
}
The reflection and transmission coefficients are
$\R=|S_{++}|^2=|S_{--}|^2$ and $\T=|S_{+-}|^2=|S_{-+}|^2$.
The normalization condition \norm\ is verified in the appendix.
(When $\kappa^2$ is too small to allow transmission from
phase 1 to phase 2, then $\R=1$, $\T=0$, and the $f^+_\kappa$
solution should be dropped.)

Using the eigenfunctions \fbasis, the properties of $S$, and the
action \oprplane\ of $\opr$, it is
straightforward to find that
\eqn\moor{
  \sum_{a,b} \left|
      {1\over2} {f_\kappa^a}^* \opr f_\kappa^b \bigr|_{-\infty}^{+\infty}
  \right|^2
  = {1\over2\pi^2} \left[ 2(k_1^2+k_2^2)\R + (k_1-k_2)^2 \T \right] .
}
[When $\kappa^2$ is too small to allow transmission from phase 1
to 2, this instead becomes simply $(2k_1^2)/2\pi^2$.]
Now use this, together with \parts, to rewrite the result \pie\
for the bubble wall velocity in terms of $\R$ and $\T$.  As promised, the
result is identical to the result \Psmall\ of the last section, where
I reviewed the simple picture of friction caused by reflection and
transmission at the bubble wall.

\newsec{A Toy Example}

I now want to specialize to Khlebnikov's toy example and understand
his results in terms of reflection and transmission probabilities.
Khlebnikov\refmark{\khlebnikov}
took $M^2(\phi) = M_0^2 + 2g\phi$ in the limit
$\Delta M^2 \ll M_0^2 \ll T^2$ where
$\Delta M^2 = M^2(z{=}+\infty) - M^2(z{=}-\infty)$.
Consider first the transmission term in the net wall pressure \Psmall,
and consider the contribution to the integral from momenta
$\Delta M^2 \ll p_z^2 \ll M_0^2$.  $\T$ is approximately 1 for such momenta,
and it is easy to extract the behavior of the integrand in these limits.
Taking $E^2 \sim M_0^2+\pt^2$, $\pza \sim \pzb$,
$\pza-\pzb \sim \Delta M^2/2\pz$, and $n(E) \sim T/E$,
this contribution to the
transmission term in \Psmall\ becomes
\eqn\Tapprox{
 \eqalign{
   \beta u \int {\d^2\pt\d E\over(2\pi)^3} & n(E) [n(E)+1]
      (\pza-\pzb)^2 \T \theta(\pzb^2)
   \cr & \approx
   u (\Delta M^2)^2 T {1\over 4(2\pi)^3}
   \int {\d p_z \over p_z} \int {\d^2\pt\over(\pt^2+M_0^2)^{3/2}}
   \cr & \approx
   u {(\Delta M^2)^2 T \over 32 \pi^2 M_0}
   \ln\left(M_0^2\over\Delta M^2\right) ,
 }
}
where $\pz$ has been integrated over the region
$\Delta M^2 \ll p_z^2 \ll M_0^2$,
and where I have ignored terms on the right-hand side that are not
enhanced by a logarithm.
It is easy to check that this contribution dominates (by the logarithm)
over the other regions of integration, $p_z^2 \lsim \Delta M^2$ and
$M_0^2 \lsim p_z^2$.

Now consider the reflection term in the net wall pressure \Psmall.
Generically, $\R$ falls to zero for $\pz^2 \gg \Delta M^2$, and so
one might not expect to get a logarithmic enhancement as in \Tapprox.
However, if the bubble wall were a {\it step} function, with zero width,
then $\R$ falls slowly enough at large $\pz$ that one
still gets a logarithm.  Specifically, for a step potential,
\eqn\Rstep{
      \R = \left| \pza-\pzb \over \pza+\pzb \right|^2 .
}
Considering the region $\Delta M^2 \ll p_z^2 \ll M_0^2$ as before
(so that $\pza\sim\pzb$ and $\T\sim 1$),
then
\eqn\Rapprox{
     2(\pza^2 + \pzb^2)\R \approx (\pza-\pzb)^2 \T ,
}
and one sees that the $\R$ and $\T$ terms in \Psmall\ contribute equally,
each giving \Tapprox.
If the spatial width $\delta$ of the wall is small but {\it non}-zero, then
$\R$ will behave like \Rstep\ for $\pz \ll 1/\delta$ but will fall faster
with $\pz$ than \Rstep\ for $\pz \gg 1/\delta$.  The logarithm found for
the transmission term in \Tapprox\ will then, for the reflection term,
be cut off by $1/\delta$ rather than $M_0$ if $1/\delta < M_0$.
At the level of the leading logarithm, the various cases are
\eqn\Rapprox{
 \eqalign{
   \beta u \int {\d^2\pt\d E\over(2\pi)^3} & n(E) [n(E)+1]
      2(\pza^2+\pzb^2) \R \theta(\pzb^2)
   \cr & \approx
   u {(\Delta M^2)^2 T \over 32 \pi^2 M_0}
   \cases{
     \ln\left(M_0^2\over\Delta M^2\right),
          & $M_0^2\lsim\delta^{-2}$, \cr
     \ln\left(\delta^{-2}\over\Delta M_0^2\right),
          & $\Delta M^2 \lsim \delta^{-2} \lsim M_0^2$, \cr
     0,
          & $\delta^{-2} \lsim \Delta M^2$ ,
   }
 }
}
where zero in the last case means that there is no term with a logarithmic
enhancement.
Putting the approximations \Tapprox\ and \Rapprox\ into the formula
\Psmall\ for the net pressure, the result for the wall velocity in
Khlebnikov's model is
\eqn\ukhleb{
   u = {32\pi^2 M_0\,\Delta\Veff \over (\Delta M^2)^2 T l}
         + O(u/l),
}
where
\eqn\ldef{
   l = \cases{
         2 \ln\left(M_0^2\over\Delta M^2\right),
           & $M_0^2 \lsim \delta^{-2}$, \cr
         \ln\left(M_0^2\over\Delta M^2\right)
         + \ln\left(\delta^{-2}\over\Delta M_0^2\right),
           & $\Delta M^2 \lsim \delta^{-2} \lsim M_0^2$, \cr
         \ln\left(M_0^2\over\Delta M^2\right),
           & $\delta^{-2} \lsim \Delta M^2$ .
        }
}
For a specific form of $\phicl(z)$, one could also calculate the
non-logarithmic correction to $l$ by precisely computing the
reflection and transmission coefficients.

Khlebnikov gave a more approximate analysis of this model where he
(1) accidentally missed the transmission term,\foot{
   The intended step between eqs.~(19) and (20) of
   ref.~\khlebnikov\ requires the identity
   $$ \delta(E-E_1) = {E\over pk}\delta\left(\cos\theta - {k\over2p}\right)
         + {E\over p|\cos\theta|} \delta(k) ,
   $$
   but the last term was left out.
   There is also a factor of 2 mistake going from eq. (20) to (21).
}
and (2) computed the contribution of \figd b to the $\PiR$ in the
approximation that the background field $\phi$ is ignored, so that the
$\Phi$ propagator is simply $1/(P^2-M_0^2)$.  This approximation turns
out to be equivalent to taking $\Delta M^2 \rightarrow 0$ in the
logarithms \ldef\ and produces a logarithmic infrared divergence.
He was forced to cut off this divergence by including rescattering
effects, which come from resumming higher order of the loop expansion.
However, we see now that there is indeed no logarithmic divergence in
the simple one-loop result.

In the case of the electroweak phase transition, $M_0$ and $\Delta M^2$
are not separate scales, and so no logarithmic enhancement of the friction
should be expected.  In electroweak models, the wall width is typically large
compared to inverse particle masses, and then
the reflection and transmission coefficients can
be well approximated by simple theta functions, $\theta(-\pzb^2)$
and $\theta(\pzb^2)$, representing complete transmission when energetically
allowed.  This approximation was used in refs.~\refs{\liu,\dinea}.

I have shown that the one-loop application of Khlebnikov's more general
fluctuation-dissipation result is equivalent to the simple physical
picture of computing the friction on the wall from particle collisions
in the ideal gas limit.
For the electroweak phase transition, however, one generally may
{\it not} ignore the effects of rescattering, and the reader should
consult refs.~\refs{\dinea-\khlebnikov} for attempts to
analyze rescattering effects.

\vskip 0.2in

I thank Lowell Brown for many long and useful discussions,
for reminding me how to do quantum mechanics, and for getting me
interested in this problem.  I also thank Larry Yaffe and Sergei
Khlebnikov for useful discussions.

\appendix{A}{}

In this appendix, I check that the solutions \fbasis\ are normalized
according to \norm.  For a complete set of states, this normalization
condition is equivalent to
\eqn\normb{
   \int \d\kappa^2 {f_\kappa^a}^*(z) f_\kappa^a(z')
   = \delta(z-z') .
}
The normalization of the basis \fbasis\ may be checked by verifying \normb\
for any particular range of $z$.
Let's focus then on $z, z' \rightarrow -\infty$.
Using \fbasis\ and the properties of $S$, one finds
\eqn\normc{
  {f_\kappa^a}^*(z) f_\kappa^a(z') \rightarrow
  {1\over4\pi k_1} \left[ e^{-ik_1(z-z')} + e^{ik_1(z-z')}
                  + 2 S_{++} e^{-ik_1(z+z')} + 2 S_{++}^* e^{ik_1(z+z')}
               \right] .
}
Then, since $\kappa^2 \rightarrow k_1^2$,
\eqn\normd{
  \int \d\kappa^2 {f_\kappa^a}^*(z) f_\kappa^a(z')
  \rightarrow \delta(z-z')
    + {1\over\pi} \int\nolimits_0^\infty \d k_1
       [S_{++} + S_{++}^*] e^{-ik_1(z+z')} .
}
The last term vanishes due to two analytic properties of $S_{++}$:
(1) $S_{++}^*(k_1) = S_{++}(-k_1)$, and (2) $S_{++}(k_1)$ is
analytic in the upper half-plane provided the potential has no
bound states (which, in applications of interest, is generally
the case).  The last term can therefore be written as
an integral from $-\infty$ to $+\infty$, which vanishes by closing
the contour in the upper-half plane.
The first property follows from the fact that, if
$f_\kappa^+(z)$ is a solution, then ${f_\kappa^+}^*(z)$ is also a solution.
For insight into the second property, suppose that $S_{++}(k_1)$ has a
{\it pole} in the upper-half plane.  At this pole,
$k_2 = \sqrt{k_1^2 - \Delta M^2}$ will
also be in the upper-half plane.  Now renormalize the solution \fbasis\
for $f_\kappa^+$
by dividing it by $S_{++}$, and then take $k_1$ to be the position of
the pole so that $S_{++}\rightarrow\infty$.  The ``incident'' wave
term disappears.  The solution falls exponentially as
$x\rightarrow\pm\infty$ and therefore describes a bound-state, which
contradicts the assumption that there are none.
(If there are bound states, the contribution from the poles of $S_{++}$
reflects the fact that the sum over bound states should have been included
on the left side of \normb.)
The simple argument above doesn't prove that there aren't other types
of singularities in the upper-half plane, and the reader should try
ref.~\scatter\ for a more general discussion of
the analytic properties of the S matrix.

\listrefs
\if\preprint0 \listfigs \fi

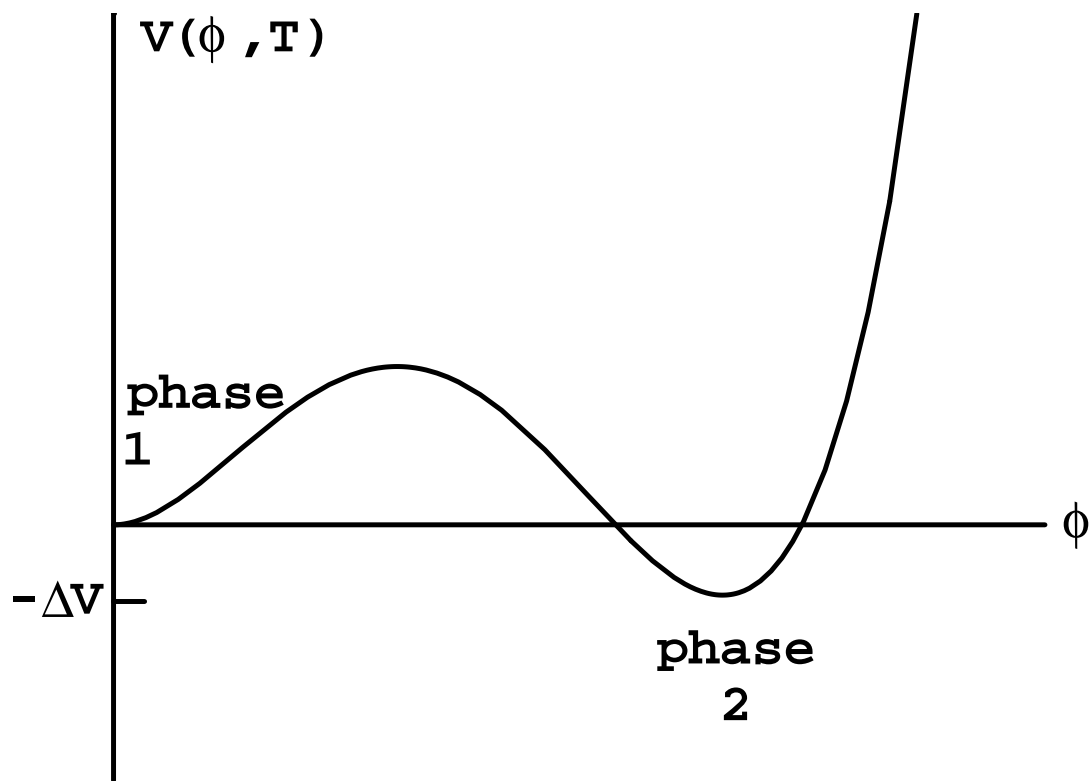
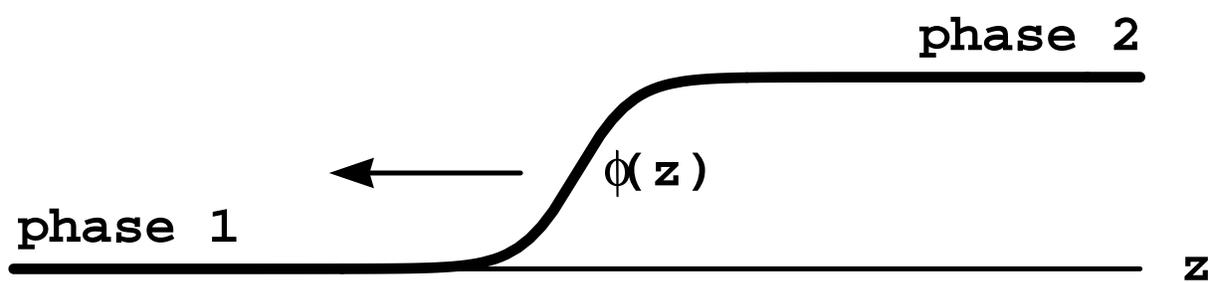
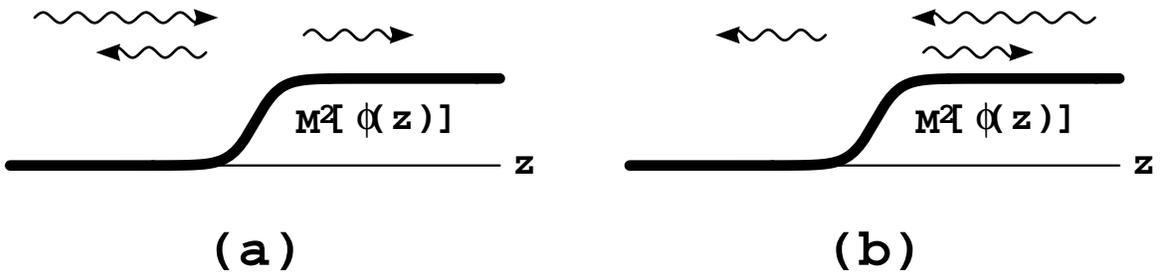
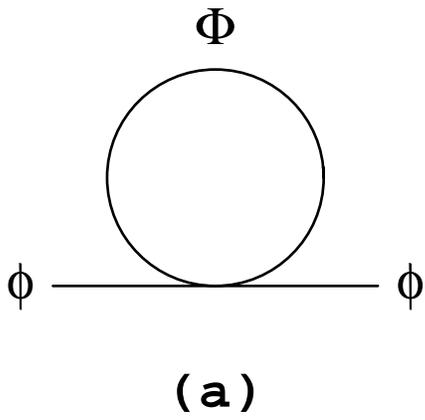
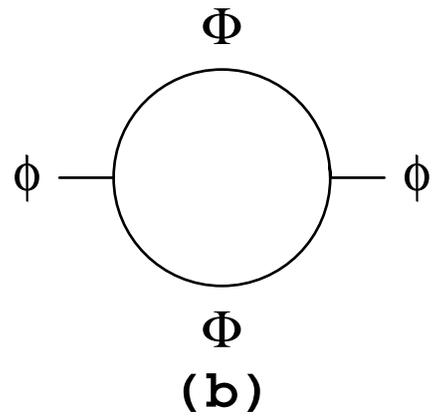
\end